\documentclass[a4paper,twocolumn,10pt,accepted=2026-03-03]{quantumarticle}
\pdfoutput=1
\usepackage[utf8]{inputenc}
\usepackage[english]{babel}
\usepackage[T1]{fontenc}
\usepackage{amsmath}
\usepackage{hyperref}
\usepackage{amssymb}
\usepackage{subcaption}
\usepackage{float}
\usepackage{tikz}
\usepackage{lipsum}
\usepackage{dsfont}

\usepackage{graphicx}
\usepackage{dcolumn}
\usepackage{bm}

\usepackage{xcolor}
\usepackage{physics}
\usepackage[normalem]{ulem}
\usepackage[numbers,sort&compress]{natbib}

\newtheorem{definition}{Definition}

\begin{document}

\title{Near-optimal coherent state discrimination via continuously labelled non-Gaussian measurements}

\author{James Moran}
\affiliation{Quantum Universe Center, Korea Institute for Advanced Study, Seoul 02455, Republic of Korea}
\email{jamesmoran@kias.re.kr}
\orcid{0000-0002-9600-9629}
\author{Spiros Kechrimparis}
\email{skechrimparis@gmail.com}
\affiliation{School of Computational Sciences, Korea Institute for Advanced Study, Seoul 02455, Republic of Korea}
\author{Hyukjoon Kwon}
\affiliation{Quantum Universe Center, Korea Institute for Advanced Study, Seoul 02455, Republic of Korea}
\affiliation{School of Computational Sciences, Korea Institute for Advanced Study, Seoul 02455, Republic of Korea}
\email{hjkwon@kias.re.kr}
\orcid{0000-0003-1985-4623}
\maketitle

\begin{abstract}
Quantum state discrimination plays a central role in quantum information and communication. For the discrimination of optical quantum states, the two most widely adopted measurement techniques are photon detection, which produces discrete outcomes, and homodyne detection, which produces continuous outcomes. While various protocols using photon detection have been proposed for optimal and near-optimal discrimination between two coherent states, homodyne detection is known to have higher error rates, with its minimum achievable error rate often referred to as the Gaussian limit. In this work, we demonstrate that, despite the fundamental differences between discretely labelled and continuously labelled measurements, continuously labelled non-Gaussian measurements can also achieve near-optimal coherent state discrimination. We design two discrimination protocols that surpass the Gaussian limit: one using non-Gaussian unitary operations with homodyne detection, and another based on orthogonal polynomials. Our results show that photon detection is not required for near-optimal coherent state discrimination and that we can achieve error rates close to the Helstrom bound at low energies with continuously labelled measurements. We also find that our schemes maintain an advantage over the photon detection-based Kennedy receiver for a moderate range of coherent state amplitudes.
\end{abstract}

\section{\label{sec:level1}Introduction}

Quantum state discrimination is a quantum information primitive and is ubiquitous in quantum communications \cite{Barnett:09,Bae_2015,9781107002173,Pirandola:20,https://doi.org/10.1049/qtc2.12015}. When two quantum states are non-orthogonal, no measurement can perfectly distinguish between them. This fundamental quantum limit is known as the Helstrom bound \cite{Helstrom:1969aa}.

In continuous-variable systems, such as quantum optical systems, quantum state discrimination is a fundamental task in optical communications \cite{Pirandola:20,https://doi.org/10.1049/qtc2.12015}. Due to their experimental availability, robustness to losses, and ease of control, the discrimination of two optical coherent states, is the simplest non-trivial example of optical state discrimination and is central to optical binary phase-shift keying protocols \cite{53264}.  One issue that arises in continuous-variable systems is that the discrimination task involves determining how to construct an optimal measurement protocol that achieves the Helstrom bound with experimentally feasible technologies. In lieu of Helstrom measurements, which are often challenging to realise, and owing to the non-uniqueness of optimal measurement setups, many receivers designed to discriminate two coherent states have appeared in the literature \cite{PhysRevLett.101.210501,PhysRevA.78.022320,Han_2018,Sidhu2023linearoptics,warke2024photonicquantumreceiverattaining}. These receivers are typically based on experiments that can be implemented with current technologies, such as homodyne detection and on-off photon detection. In the literature thus far, all known receivers that surpass the homodyne limit, with some even approaching the Helstrom bound, rely on a final on-off photon detection stage. The sole exception is a scheme involving a Kerr gate followed by homodyne detection, which was also shown to exceed the homodyne limit \cite{PhysRevA.51.1702,Sasaki-Usuda:1995aa}.

It was shown in Ref.~\cite{PhysRevA.78.022320} that achieving the Helstrom bound for discriminating two coherent states requires non-Gaussian measurements. Specifically, while the optimal performance among Gaussian measurement schemes—known as the \emph{Gaussian limit}—is attained by homodyne detection, which measures a quadrature operator, an exponential gap in error scaling remains between the Gaussian limit and the Helstrom bound. Receivers that are able to achieve the Helstrom bound appeared in Refs.~\cite{dolinarRe, PhysRevA.54.2728} which are known as the Dolinar and Sasaki-Hirota receivers, respectively. In both cases, on-off photon detection, a non-Gaussian operation, is used, and in the Sasaki-Hirota receiver, an additional non-Gaussian unitary is applied. This raises several questions: Is photon detection necessary to achieve the Helstrom bound in coherent state discrimination? If we are restricted to continuously labelled measurement schemes, such as those based on homodyne or heterodyne detection, by how much can we surpass the Gaussian limit in approaching the Helstrom bound? Moreover, what is the form of the non-Gaussian operations that enable this improvement?

In this work, we address these questions: we define two types of continuously labelled receivers that can exceed the Gaussian limit. The first is constructed from a class of non-Gaussian unitaries which, when combined with homodyne detection, lead to significant performance improvements over the Gaussian limit in both low- and high-energy regimes. The second type is based on the theory of orthogonal polynomials. We find that generalised states based on Legendre and Laguerre polynomials can also exceed the Gaussian limit. This confirms that photon detection is not required for significant improvement in performance beyond the Gaussian limit in coherent state discrimination, and this regime can be achieved with continuously labelled measurements. However, the results do not specify the precise form of non-Gaussianity required. To this end we introduce a quantification of the non-Gaussianity of measurements. We also provide examples of continuously labelled non-Gaussian measurement schemes with varying degrees of non-Gaussianity, which are shown to degrade performance below the Gaussian limit.

The paper is structured as follows. In Sec.~\ref{CohStaDisc} we recapitulate the problem of minimum-error quantum state discrimination, the notion of Gaussian and non-Gaussian states and measurements, and the issues that arise when the states to be discriminated belong to infinite-dimensional Hilbert spaces. In Sec.~\ref{WavPar} we introduce the notion of continuously and discretely labelled measurements and analyse their fundamental differences and connection to wave-particle duality. Following this, in Sec.~\ref{ngrotsec} we define a first type of continuously labelled receivers based on non-Gaussian unitaries combined with Gaussian measurements and show that they exceed the Gaussian limit. In Sec.~\ref{LegPolSec}, we define a second type of continuously labelled receiver based on the orthogonality of the Legendre polynomials, and show that this receiver also surpasses the Gaussian limit. In Sec.~\ref{StelRan} we discuss the role of non-Gaussianity in coherent state discrimination, and define a measure of non-Gaussianity of quantum measurements based on the stellar rank. We find two examples at finite and infinite non-Gaussianity which are detrimental to the error rate, showing that non-Gaussianity does not guarantee improved discrimination performance. Lastly, in Sec.~\ref{conclusion}, we conclude by summarising our findings and addressing future directions this work could take.

\section{Coherent state discrimination}\label{CohStaDisc}
\subsection{Minimum-error discrimination}\label{MinErrDisc}
Minimum-error discrimination of two states, $\hat{\rho}_1, \hat{\rho}_2$, with prior probabilities, $p_1, p_2=1-p_1$, respectively, is characterised by a two-element positive operator-valued measure (POVM), $\mathcal{M}=\left\{\hat{\Pi}_1, \hat{\Pi}_2=\mathbb{I}-\hat{\Pi}_1\right\}$, where $\hat{\Pi}_i$ is the measurement operator corresponding to hypothesis $\mathcal{H}_i$, that guesses the state's label as $i$. For non-orthogonal quantum states there is a  non-zero error in determining which state was sent. Concretely, the probability of error, $P_\textrm{E}$, follows from Bayes' probability law \cite{Barnett:09,Bae_2015}
\begin{equation}
\label{BornErr}
 P_\textrm{E}=p_1 \tr\left[\hat{\Pi}_2\hat\rho_1 \right]+p_2\tr \left[\hat{\Pi}_1\hat\rho_2 \right].
\end{equation}

The minimisation of Eq.~\eqref{BornErr}, with respect to $\hat{\Pi}_i$, leads to the Helstrom bound \cite{Helstrom:1969aa}, which describes the fundamental quantum limit on the minimum-error discrimination of two non-orthogonal quantum states. The Helstrom bound is given by
\begin{equation}\label{HelBound}
P_{\textrm{H}}=\frac{1}{2}\left(1-\tr \| p_1 \hat{\rho}_1 -p_2 \hat{\rho}_2 \|_1 \right),
\end{equation}
where $\| \hat{O} \|_1=\tr\left[\sqrt{\hat{O}^\dagger \hat{O}}\right]$ denotes the trace-norm. 

In this work we will restrict our attention to the case of two pure states, $\hat{\rho}_i=\ket{\psi_i}\bra{\psi_i}$, and equal prior probabilities, $p_1=p_2=\frac{1}{2}$. The Helstrom bound Eq.~\eqref{HelBound} becomes
\begin{equation}
 P_\textrm{H}=\frac{1}{2}\left(1-\sqrt{1-\abs{\bra{\psi_1}\ket{\psi_2}}^2}\right),
\end{equation}
and the optimal measurements obtained from the theory are projections onto the positive and negative eigenspaces of the operator $\delta\hat{\rho}=\hat{\rho}_1-\hat{\rho}_2$.
\subsection{Coherent states and Gaussian measurements}
A class of states ubiquitous in quantum optics and continuous-variable systems are the Glauber-Sudarshan coherent states \cite{PhysRev.130.2529,PhysRevLett.10.277}. Coherent states are often regarded as the most classical of quantum states, as they minimise the Heisenberg uncertainty relation and their semiclassical equations of motion closely resemble those of classical harmonic oscillators. The Fock state expansion of a coherent state is given by:
\begin{equation}\label{CohStat}
 \ket{\alpha}=e^{-\frac{\abs{\alpha}^2}{2}}\sum_{n=0}^\infty \frac{\alpha^n}{\sqrt{n!}}\ket{n}, \quad \alpha \in \mathbb{C}.
\end{equation}

Coherent states may also be represented by a complex displacement of the vacuum state in phase space by the displacement operator
\begin{equation}\label{DispOp}
 \hat{D}(\alpha)=\exp\left(\alpha \hat{a}^\dagger - \bar{\alpha}\hat{a}\right),
\end{equation}
such that $\ket{\alpha}= \hat{D}(\alpha)\ket{0}$.
Coherent states are referred to as \emph{Gaussian states} because their Wigner phase space distributions are Gaussian \cite{RevModPhys.84.621}. Gaussian operations are then defined as operations that transform Gaussian states to Gaussian states, and Gaussian unitaries may be formed by the unitary evolution of Hamiltonians which are at most quadratic in the creation and annihilation operators \cite{RevModPhys.84.621,doi:10.1142/S1230161214400010,PRXQuantum.2.030204}.

This also underpins the framework for Gaussian measurements.
Gaussian measurements are described by the following POVM elements \cite{RevModPhys.84.621,PhysRevA.66.032316}
\begin{equation}\label{GausMeas}
 \mathrm{d}^2\beta\, \hat{\Pi}_{\textrm{G}}(\beta)= \mathrm{d}^2\beta\, \frac{1}{\pi}\hat{D}(\beta) \hat{\rho}_{\textrm{G}}\hat{D}^\dagger(\beta),
\end{equation}
satisfying the normalisation condition
\begin{equation}\label{IntReg}
 \int_{\beta \in\mathbb{C}}\mathrm{d}^2\beta\,\hat{\Pi}_{\textrm{G}}(\beta)=\mathbb{I}.
\end{equation}
Here, $\hat{\rho}_{\textrm{G}}$ denotes a zero-mean Gaussian state. Gaussian states are fully characterised by their covariance matrix and coherent displacement vector; in the zero-mean case, the displacement vector is zero. In this setting, a two-outcome POVM as described in Sec.~\ref{MinErrDisc} would be defined by splitting the integration region \eqref{IntReg} into two distinct regions, $\mathcal{R}_1, \mathcal{R}_2$, satisfying $\mathcal{R}_1 \cup \mathcal{R}_2=\mathbb{C}$ and $\mathcal{R}_1 \cap \mathcal{R}_2=\emptyset $, with the additional requirement that both $\mathcal{R}_1$ and $\mathcal{R}_2$ have non-zero Lebesgue measure. Gaussian measurements include homodyne detection (i.e. projections onto quadrature eigenstates), which may be interpreted through \eqref{GausMeas} as a measurement of infinitely squeezed coherent states, and heterodyne detection (i.e. projections onto coherent states) \cite{djord,wallsmilburn}.

An identity related to the total variation distance between two probability distributions, which will be useful later, is given by \cite{PhysRevA.103.022423}:
\begin{equation}\label{totvardist}
 P_\textrm{E}=\frac{1}{2}-\frac{1}{4}\int_{x\in\mathcal{X}}\mathrm{d}x\, \abs{\rho_1(x)-\rho_2(x)},
\end{equation}
where $\rho_i (x)=\tr\left[\hat\Pi_x \hat \rho_i\right]$ is the continuous probability distribution induced by the POVM $\left\{\mathrm{d}x\,\hat\Pi_x\right\}$. This formula allows us to replace the integration over a to-be-determined region, $\mathcal{R}_2$ (the region where $ \left(\rho_1(x)-\rho_2(x)\right)$ is negative), by an integration of the absolute value of the difference between the distributions over the entire space of outcomes.

\subsection{Coherent state discrimination}
Applying the theory of minimum-error discrimination to two coherent states with opposite signs in their amplitudes, $\left\{\ket{\alpha},\ket{-\alpha}\right\}, \alpha\in\mathbb{R}$, yields an optimal measurement of the form \cite{Helstrom:1969aa}
\begin{equation}\label{CatMeas}
 \hat{\Pi}_2=\ket{-}\bra{-},
\end{equation}
where
\begin{equation}
\begin{split}
 \ket{-}=&\left(\frac{1-\sqrt{1-e^{-4\alpha^2}}}{\sqrt{1-e^{-4\alpha^2}}}\right)\ket{\alpha}\\
 &+\left(2e^{-2\alpha^2}-\frac{e^{-2\alpha^2}}{\sqrt{1-e^{-4\alpha^2}}}\right)\ket{-\alpha},
 \end{split}
\end{equation}
and $\hat{\Pi}_1=\ket{+}\bra{+}=\mathbb{I}-\ket{-}\bra{-}$, with the identity operator here being valid on the signal space, $\mathbb{I}\ket{\pm\alpha}=\ket{\pm\alpha}$.

The optimal measurement in state discrimination problems is not unique \cite{Barnett:09,Bae_2015}. Moreover, the actual implementation of the Helstrom measurement of the form of Eq.~\eqref{CatMeas} is a projection onto a cat state, and is impractical with today's technologies. Only recently have projective measurements onto cat states been performed \cite{Izumi:2018aa}, but this is only in the low-amplitude limit and it is achieved with poor fidelity due to detector imperfections. With these limitations in mind, various optical receivers designed to approach and in some cases saturate the Helstrom bound for coherent state discrimination have been conceived.

The Helstrom bound for discriminating two coherent states is
\begin{equation}
 P_{\textrm{E}}^{\textrm{Helstrom}}=\frac{1}{2}\left(1-\sqrt{1-e^{-4\alpha^2}}\right),
\end{equation}
while the limit when restricted to Gaussian measurement schemes is given by \cite{PhysRevA.78.022320}
\begin{equation}
 P_{\textrm{E}}^{\textrm{G}}=\frac{1}{2}\left(1- \erf\left(\sqrt{2}\abs{\alpha}\right)\right),
\end{equation}
and the Gaussian limit is achieved by homodyne detection. We see that there is a significant gap between these two functions at finite non-zero $\alpha$. This highlights the importance of quantum measurements in state discrimination scenarios: if we are able to go beyond the Gaussian limit, we may see potentially exponential improvements in error rates.

From here on we will refer to any receiver whose performance lies strictly between the Gaussian limit and the Helstrom bound for any range of $\abs{\alpha}^2$ as \emph{near-optimal}. We will also provide examples of receivers which do not achieve optimality or near-optimality despite invoking non-Gaussian operations.

Given that the Helstrom measurement, Eq.~\eqref{CatMeas}, may be implemented through photon detection \cite{PhysRevA.54.2728}, almost all optimal and near-optimal receivers in the literature make use of photon detection \eqref{OnOff} to surpass the Gaussian limit. The most notable near-optimal example is the Kennedy receiver \cite{kennedy} which uses displacement followed by on-off photon detection. The measurement setup is defined by the POVM $\mathcal{M}_\textrm{K}=\left\{\hat{\Pi}_1, \hat{\Pi}_2\right\}$, with measurement operators
\begin{equation}\label{KenRec}
 \hat{\Pi}_1 = \hat{D}^\dagger(\alpha)\hat{\Pi}_{\textrm{off}}\hat{D}(\alpha), \quad  \hat{\Pi}_2 = \hat{D}^\dagger(\alpha)\hat{\Pi}_{\textrm{on}}\hat{D}(\alpha),
\end{equation}
where
\begin{equation}\label{OnOff}
 \hat{\Pi}_\textrm{off}=\ket{0}\bra{0}, \quad \hat{\Pi}_\textrm{on}=\sum_{n=1}^\infty \ket{n}\bra{n}.
\end{equation}
The principal idea is to displace the alphabet $\left\{\ket{\alpha},\ket{-\alpha}\right\}\rightarrow \left\{\ket{0},\ket{2\alpha}\right\}$ and then perform on-off detection, where a measurement outcome of `on' would indicate that the state is $\ket{2\alpha}$, and an outcome of `off' would be ambiguous, but there is higher probability that it is the state $\ket{0}$. In the high-energy regime, the Kennedy receiver gives lower error rates than the Gaussian limit and has the same error exponent as the Helstrom bound.

The first optimal receiver to appear in the literature was the Dolinar receiver \cite{dolinarRe}. The Dolinar receiver expands on the Kennedy receiver by including real-time feedback such that the displacement parameter in \eqref{KenRec} is dynamically updated according to the intermediate measurement outcomes. The Dolinar receiver was also shown to be robust in the presence of noise \cite{PhysRevA.70.062303}, and it may be interpreted as a multiple-copy discrimination problem \cite{PhysRevA.84.022342}. The Dolinar principle has also been demonstrated experimentally \cite{Cook:2007aa}, but continuous feedback techniques come with several drawbacks, including low symbol repetition rate and the need for fast detectors.

The Sasaki-Hirota receiver was also shown to be optimal for coherent state discrimination and it simplifies the need for a complex feedback mechanism at the expense of requiring a non-Gaussian unitary rotation onto a cat state as a preprocessing, followed by on-off photon detection. Unlike the Dolinar receiver, the Sasaki-Hirota receiver was shown to not be robust in the presence of noise.

A variety of other near-optimal receivers have been proposed in the literature for coherent state discrimination and related problems \cite{53264,60560,Muller:2012aa}, typically based on operations that are feasible with current technologies. In almost all known optimal and near-optimal receivers, photon detection is used. Motivated by this, we investigate the utility of continuously labelled measurements—i.e., those that exclude photon detection—for coherent state discrimination.

\section{Continuously and discretely labelled measurements}\label{WavPar}
Homodyne detection and photon detection correspond to measurements of fundamentally different aspects of a quantum system: the former probes wave-like properties, while the latter reveals particle-like behaviour. This distinction reflects the broader phenomenon of wave–particle duality, wherein quantum systems exhibit both wave-like and particle-like characteristics \cite{PhysRevLett.77.2154, Aiello2023probabilisticviewof}. In the context of the problem explored in this paper, this duality manifests as a fundamental inequivalence between measurements with continuous and discrete spectra. For a review on discrete- and continuous-variable quantum information, we refer the reader to \cite{RevModPhys.77.513}.

To see this inequivalence explicitly, we show that continuously labelled measurements cannot reproduce on-off photon counting in Eq.~\eqref{OnOff}. Consider the following family of continuously labelled operators, $\ket{\psi}\bra{\psi}$, satisfying a completeness relation,
\begin{equation}\label{ContDiscDiff}
 \int_{\psi \in \mathcal{X}}\mathrm{d}\psi\, \mu\left(\psi\right) \ket{\psi}\bra{\psi}=\sum_{n=0}^\infty \ket{n}\bra{n}=\mathbb{I},
\end{equation}
for some positive integral measure $\mu\left(\psi\right)$ on some uncountably infinite set $\mathcal{X}$. If on-off detection were possible by a measurement on $\psi$, then there should exist a region $\mathcal{Y} \subset \mathcal{X}$, with $\mathcal{Y}$ being dense $\mathcal{X}$, such that
\begin{equation}\label{ContDiscDiffSub}
 \int_{\psi \in \mathcal{Y}}\mathrm{d}\psi\, \mu\left(\psi\right) \ket{\psi}\bra{\psi}= \hat{\Pi}_\textrm{off}.
\end{equation}
The expansion of $\ket{\psi}$ in the number basis is
\begin{equation}\label{psiExp}
 \ket{\psi}=\sum_{n=0}^\infty c_n \left(\psi\right) \ket{n},
\end{equation}
where we need not impose any normalisation restriction on the coefficients $c_n \left(\psi\right)$, so that this setup accounts for measurements like homodyne detection. However, we observe that $\forall n$ there must exist some range of $\psi \in \mathcal{Z}_n$ where
\begin{equation}
 \mathcal{Z}_n=\left\{\psi \in \mathcal{X}\mid c_n(\psi)\not=0 \right\},
\end{equation}
with $\mathcal{Z}_n \subseteq \mathcal{X}$, dense in $\mathcal{X}$. This condition follows from the requirement that we have non-zero projectors onto every number state in \eqref{ContDiscDiff}, and thus each coefficient $c_n(\psi)$ must be non-zero at least somewhere in $\mathcal{X}$. While we do not impose a normalisation condition on these coefficients, they are subject to orthogonality constraints, from which the following integral identities can be derived:
\begin{equation}
  \int_{\psi \in \mathcal{X}}\mathrm{d}\psi\, \mu\left(\psi\right) c_n\left(\psi\right) \bar{c}_m\left(\psi\right)=\delta_{nm}, \quad \forall n,m.
\end{equation}

Expanding Eq.~\eqref{ContDiscDiffSub} in the number basis yields the following matrix elements:
\begin{equation}
 \mathcal{I}_{nm}=  \int_{\psi \in \mathcal{Y}}\mathrm{d}\psi\, \mu\left(\psi\right) c_n\left(\psi\right) \bar{c}_m\left(\psi\right)\ket{n}\bra{m},
\end{equation}
with the diagonal terms given by:
\begin{equation}
 \mathcal{I}_{nn}=  \int_{\psi \in \mathcal{Y}}\mathrm{d}\psi\, \mu\left(\psi\right) \abs{c_n\left(\psi\right)}^2 \ket{n}\bra{n}.
\end{equation}
Since the coefficients $\abs{c_n(\psi)}^2$ are non-negative for all $n$ and $\psi$, we have the bound
\begin{equation}
0 \leq \bra{n} \mathcal{I}_{nn} \ket{n} \leq 1,
\end{equation}
with the upper bound saturated only when $\mathcal{Y} = \mathcal{X}$. The implication of this is that the only way to achieve exact projection onto the ground state is by taking the set $\mathcal{Y}=\mathcal{X}$, but this of course does not realise photon-counting as we have projections onto all diagonal states with coefficient $1$, i.e., we have reproduced the identity operator \eqref{ContDiscDiff}. Thus, there is no way to exactly reproduce on-off photon counting by a continuously labelled measurement of the form of Eq.~\eqref{ContDiscDiffSub}. This is shown explicitly for the examples of heterodyne and homodyne detection in App \ref{HomIneq}. We remark that we demand equivalence in the strict sense (i.e. an equality). One may attempt to construct a continuously labelled measurement, which up to an error can reproduce a vacuum state projection, but we exclude such examples in the present work. We are also excluding reconstructions of photon-number statistics through phase-averaged homodyne measurements \cite{PhysRevA.52.R924}, as we are focusing on single-shot measurement schemes. For a comparison against near-optimal single-shot schemes using photon detection, we refer the reader to App.~\ref{compapp}.

Another important remark is that photon-counting is \emph{always} non-Gaussian, while certain types of continuously labelled non-Gaussian measurements may be interpreted as Gaussian measurement of a non-Gaussian-transformed state. These examples include preprocessing by non-Gaussian unitaries followed by homodyne detection. We may interpret this as a homodyne (Gaussian) detection of a preprocessed non-Gaussian state. We will consider examples of this in the following.

\section{Non-Gaussian unitaries combined with homodyne measurement}\label{ngrotsec}
\begin{figure}\center
\includegraphics[scale=0.33]{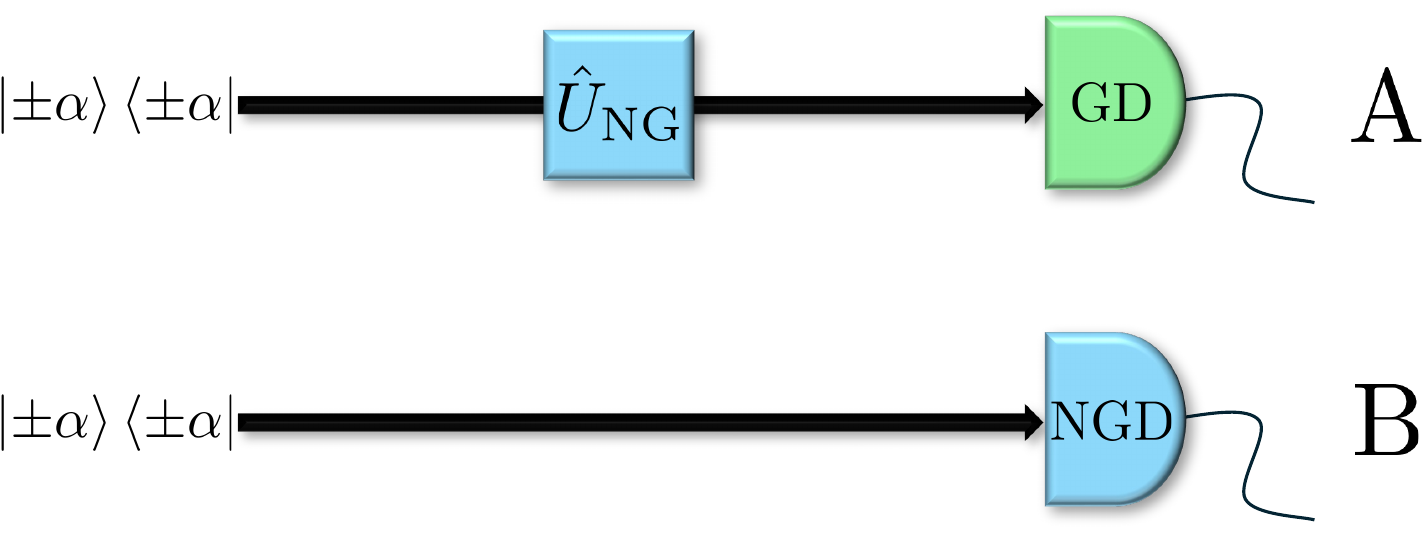}
\caption{\label{type1type2}A pictorial representation of the two types of continuously labelled non-Gaussian measurement schemes considered in this work. Type A refers to a non-Gaussian unitary preprocessing of the coherent states followed by Gaussian detection. Type B refers to a continuously labelled non-Gaussian measurement which is unitarily inequivalent to type A.}
\end{figure}

One straightforward way to construct continuously labelled non-Gaussian measurements is to apply non-Gaussian unitary operations followed by Gaussian measurements. This leads to the following definition:
\begin{definition}\label{type1}
Type A receivers are defined by a unitary preprocessing of the states, followed by Gaussian measurements. The POVM elements are given by
\begin{equation}\label{t1povm}
\mathrm{d}^2 \beta\,\hat{\Pi}_{\emph{\textrm{A}}}=\mathrm{d}^2 \beta\,\hat{U}^\dagger_{\emph{\textrm{NG}}} \hat{\Pi}_{\emph{\textrm{G}}}(\beta)\hat{U}_{\emph{\textrm{NG}}} .
\end{equation}
\end{definition}
Here $ \hat{\Pi}_{\textrm{G}}(\beta)$ is the Gaussian POVM from Eq.~\eqref{GausMeas}, and $\hat{U}_{\textrm{NG}}$ is some non-Gaussian unitary. This scheme is depicted in Fig.~\ref{type1type2}.

Given that homodyne detection defines the Gaussian limit, we look at the subset of type A measurements where the Gaussian detection stage is homodyne detection,
\begin{equation}
\mathrm{d}x\, \hat{\Pi}_{\textrm{A}}= \mathrm{d}x\, \hat U_{\textrm{NG}}^\dagger \ket{x}\bra{x}\hat U_{\textrm{NG}}.
\end{equation}
In our setup, this amounts to a preprocessing our coherent states by
\begin{equation}
 \ket{\pm \alpha} \rightarrow \hat U_{\textrm{NG}} \ket{\pm \alpha},
\end{equation}
and then performing homodyne detection on the resulting non-Gaussian states. We are thus looking for non-Gaussian operations that lead to two distributions with increased total variation distance by Eq.~\eqref{totvardist}.
We remark that the first example of such a receiver in coherent state discrimination was described in \cite{PhysRevA.51.1702,Sasaki-Usuda:1995aa}, where a non-Gaussian Kerr unitary, $\hat{U}_{\textrm{Kerr}}=\exp\left(\mathrm{i}\kappa\hat{a}^{\dagger 2}\hat{a}^2\right)$, $\kappa \in \mathbb{R}$, is applied to the coherent state before homodyne detection. For a comparison against our schemes, we refer the reader to App.~\ref{IncN}.
\begin{figure*}\center
\includegraphics{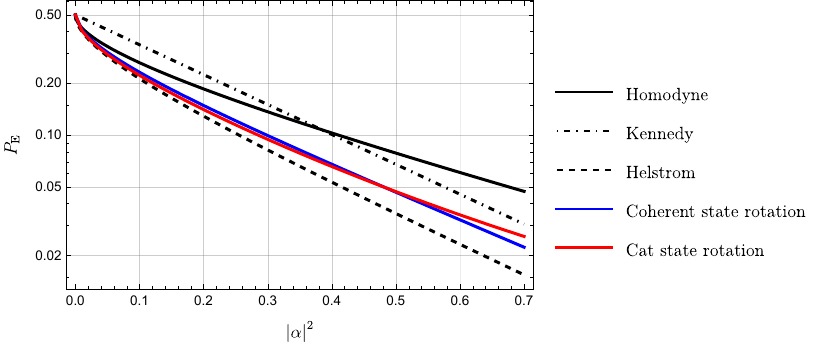}
\caption{\label{lowenergy}Comparison of the two best performing near-optimal continuously labelled receivers in Tab.~\ref{POVMtab} versus the homodyne, Kennedy and Helstrom limits in the energy range $0<\abs{\alpha}^2<0.7$. Here we find that the cat state rotation and coherent state rotation perform best}
\end{figure*}
\subsection{Unitary rotations by pure state projectors}\label{UnitSec}
As a specific example of a non-Gaussian unitary operation, we consider the following family of rotations:
\begin{equation}\label{NGrotU}
 \hat{U}^{(N)}_{\left\{\ket{\psi_k}\right\}}\left(\vec{\theta}\right)=\prod_{k=0}^N \exp\left[-\mathrm{i} \theta_k  \ket{\psi_k}\bra{\psi_k}\right],
\end{equation}
where $\vec{\theta}=\left( \theta_0, \theta_1, \ldots,  \theta_N\right)$ is an $(N+1)$-dimensional vector. These unitary gates represent a generalisation of the selective number-dependent arbitrary phase (SNAP) gate, which is realised by letting $\left\{\ket{\psi_k}\right\}=\left\{\ket{k}\right\}$, \cite{PhysRevLett.115.137002, Park_2024}. SNAP gates are realisable on qubit-oscillator platforms with existing technologies, and they provide an efficient method for simulating non-linear dynamics which are necessary for universal quantum information processing with continuous variables \cite{RevModPhys.77.513}.

With the correct choice of states, $\left\{\ket{\psi_k}\right\}$, the unitary rotations in Eq.~\eqref{NGrotU} allow for near-optimal coherent state discrimination. When combined with a homodyne detector, the final POVM elements are given by
\begin{equation}\label{UFrec1}
\mathrm{d}x\, \hat{\Pi}^{(N)}_{\left\{\ket{\psi_k}\right\}}\left(\vec{\theta}\right)= \mathrm{d}x\, \left[ \hat{U}^{(N)}_{\left\{\ket{\psi_k}\right\}}\left(\vec{\theta}\right)\right]^\dagger\ket{x}\bra{x} \hat{U}^{(N)}_{\left\{\ket{\psi_k}\right\}}\left(\vec{\theta}\right).
\end{equation}
The action of these unitary operations on an arbitrary input state is 
\begin{equation}\label{NGrotstat1}
 \hat{U}^{(N)}_{\left\{\ket{\psi_k}\right\}}\left(\vec{\theta}\right)\ket{\Psi}= \ket{\Psi}+\sum_{k=0}^N\left(\sum_{\sigma \in \mathcal{T}} \prod_{j=0}^k \zeta_{\sigma(j)}\right)\ket{\Psi},
\end{equation}
where
\begin{equation}\label{NGrotstat}
\begin{split}
 \sigma \in \mathcal{T} \iff& \left\{\sigma: \left\{0,\ldots,k\right\} \rightarrow \left\{0,\ldots N\right\} \textrm{is injective} \right\}\\
 & \wedge \left\{ k<N \implies \sigma(k)<\sigma(k+1)\right\},
 \end{split}
\end{equation}
with $\abs{\mathcal{T}}=\binom{N+1}{k+1}$ and
\begin{equation}
 \zeta_k=\left(e^{-\mathrm{i}\theta_k}-1\right)\ket{\psi_k}\bra{\psi_k}.
\end{equation}
The set $\mathcal{T}$ ensures that all the products of terms appear in the correctly indexed order given that the terms inside Eq.~\eqref{NGrotU} in general do not commute. When the states $\left\{\ket{\psi_k}\right\}$ are all orthogonal, Eq.~\eqref{NGrotstat1} simplifies to
\begin{equation}\label{eq:rot_state}
  \hat{U}^{(N)}_{\left\{\ket{\psi_k}\right\}}\left(\vec{\theta}\right)\ket{\Psi}=\ket{\Psi}+ \sum_{k=0}^N\left(e^{-\mathrm{i}\theta_k}-1\right)\bra{\psi_k}\ket{\Psi}\ket{\psi_k}.
\end{equation}
The optimal choice of the parameters $\vec{\theta}$ will depend on the input state. As we want to discriminate two coherent states, $\left\{\ket{\alpha},\ket{-\alpha}\right\}$, these optimal parameters will in general implicitly depend on the intensity $\abs{\alpha}^2$. We remark that while the states we rotate by, $\left\{\ket{\psi_k}\right\}$, may be Gaussian, the unitary rotation $\hat{U}^{(N)}_{\left\{\ket{\psi_k}\right\}}\left(\vec{\theta}\right)$ is in general not.

If we let $N=0$ such that we rotate by one state, the optimal rotation angle can be determined exactly, this is because if any improvement is obtained in any region by such a rotation, it is scaled by a factor proportional to $(1-\cos \theta_0)$, which is maximised at $\theta_0=\pi$. To see this, consider the rotation of a real coherent state, $\ket{\pm \alpha}$, by $\hat{U}^{(0)}_{\left\{\ket{\psi}\right\}}\left(\vec{\theta}\right)$, where $\ket{\psi}$ has a real-valued wavefunction. We find the following probability density
\begin{equation}
\begin{split}
 &\abs{\bra{x} \hat{U}^{(0)}_{\left\{\ket{\psi}\right\}}\left(\vec{\theta}\right)\ket{\alpha}}^2=\abs{\bra{x}\ket{\alpha}}^2\\
&+2(1-\cos\theta)\left(\abs{\bra{\psi}\ket{\alpha}}^2 \abs{\bra{x}\ket{\psi}}^2-\bra{x}\ket{\alpha}\bra{\psi}\ket{\alpha}\bra{x}\ket{\psi} \right)\\
&=\abs{\bra{x}\ket{\alpha}}^2+2(1-\cos \theta) \mathcal{F}(x,\alpha,\psi),
\end{split}
\end{equation}
where we have defined the quantity
\begin{equation}
 \mathcal{F}(x,\alpha,\psi)=\abs{\bra{\psi}\ket{\alpha}}^2 \abs{\bra{x}\ket{\psi}}^2-\bra{x}\ket{\alpha}\bra{\psi}\ket{\alpha}\bra{x}\ket{\psi}.
\end{equation}
When we compute the error rate, $P_\textrm{E}$, we find
\begin{equation}\label{thetaproof}
\begin{split}
 P_{\textrm{E}}&=\frac{1}{2}+\frac{1}{2}\int_{x\in\mathcal{R}_2}\mathrm{d}x\, \left(\abs{\bra{x}\ket{\alpha}}^2-\abs{\bra{x}\ket{-\alpha}}^2\right)\\
 &+(1-\cos\theta)\int_{x \in\mathcal{R}_2}\mathrm{d}x\,\left(\mathcal{F}(x,\alpha,\psi)-\mathcal{F}(x,-\alpha,\psi)\right),
\end{split}
\end{equation}
where the top line of \eqref{thetaproof} must be upper bounded by the homodyne limit, $P_{\textrm{E}}^{\textrm{G}}$, and if a non-Gaussian enhancement in the error rate occurs, it must then occur in the second line, which will always be optimised by $\theta=\pi$. If we consider $N>0$, the optimal rotation angles will in general change. For more details on increasing $N$, see App.~\ref{IncN}.

In the following subsections we will consider different sets of states $\left\{\ket{\psi_k}\right\}$, typical to quantum optics, and show that POVMs of the form \eqref{UFrec1} have a non-Gaussian advantage over homodyne detection for coherent state discrimination, with homodyne detection being recovered when $\vec{\theta}=\vec{0}$.

\subsection{Unitary cat state rotations}
Motivated by the appearance of cat states in the optimal measurements in Eq.~\eqref{CatMeas}, we observe that we can generate cat states by an appropriate choice of states in Eq.~\eqref{eq:rot_state}. Consider the following
 \begin{equation}
 \left\{\ket{\psi_k}\right\}=\left\{\ket{\textrm{cat}(\beta_k,\varphi_k)}\right\},
\end{equation}
where a general two-headed cat state is given by
\begin{equation}
 \ket{\textrm{cat}(\beta,\varphi)}=\frac{1}{\sqrt{2(1+e^{-2\abs{\beta}^2}\cos \varphi) }}\left(\ket{\beta}+e^{\mathrm{i}\varphi}\ket{-\beta}\right).
\end{equation}
We set $N = 0$ and take $\varphi = 0$, so that the unitary reduces to a rotation by a single even cat state. This generates a three-headed cat state by Eq.~\eqref{eq:rot_state}. In the energy regime $0<\abs{\alpha}^2<0.47$, we find a significant advantage over a homodyne receiver, the Kennedy receiver, and other near-optimal receivers considered listed in Tab.~\ref{POVMtab}. In the high-energy regime the performance of the cat state rotation falls off, though it maintains an advantage over homodyne detection. The performance of the cat states is plotted in Fig.~\ref{lowenergy}. For details on the optimal scaling of the cat state parameter, $\beta$, see App.~\ref{ParamApp}.

\subsection{Unitary coherent state rotations}
Next, we consider rotations generated by coherent states,
 \begin{equation}
 \left\{\ket{\psi_k}\right\}=\left\{\ket{\beta_k}\right\},
\end{equation}
which, according to Eq.~\eqref{eq:rot_state}, produce a two-headed cat state. As before, we set $N = 0$ and take the optimal rotation angle $\vec{\theta} = (\pi)$. When this unitary is applied to the input coherent states $\left\{\ket{\alpha},\ket{-\alpha}\right\}$, it effectively generates two cat states, which we then discriminate using homodyne detection. We may rewrite this gate in terms of realisable gates using the following decomposition:
\begin{equation}\label{snapdisp}
  \hat{U}^{(0)}_{\left\{\ket{\beta_0}\right\}}\left(\pi\right)=\hat{D}\left(\beta_0 \right) \hat{U}^{(0)}_{\left\{\ket{0}\right\}}(\pi)\hat{D}^{\dagger}\left(\beta_0 \right).
\end{equation}
Here $\hat{U}^{(0)}_{\left\{\ket{0}\right\}}(\pi)$ is a SNAP gate \cite{PhysRevLett.115.137002, Park_2024}, and we do some additional Gaussian optimisations through displacements by the parameter $\beta_0$. SNAP gates have already been demonstrated in cavity quantum electrodynamics, and, in particular, unitary operations of the form in Eq.~\eqref{snapdisp} have been realised \cite{PhysRevLett.115.137002}.

In the low-energy regime, we observe performance improvements comparable to those obtained using Fock state rotations. While the improvement is not as pronounced in the very low-energy limit compared to cat-state-based rotations, it remains more stable across a broader range of energies.

\begin{figure*}\center
\includegraphics{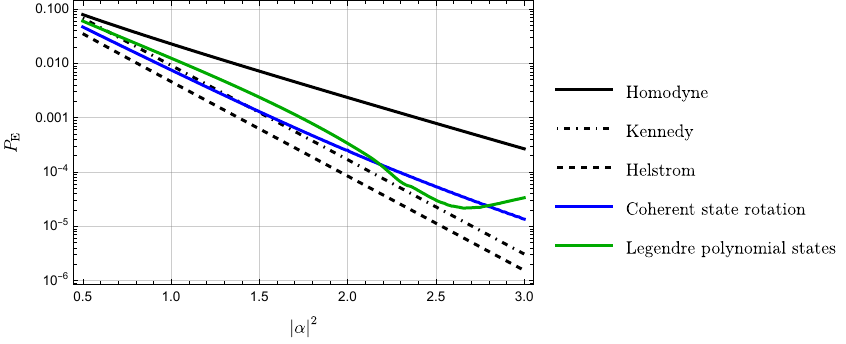}
\caption{\label{highenergy}Comparison of the two best performing near-optimal continuously labelled receivers in Tab.~\ref{POVMtab} versus the homodyne, Kennedy and Helstrom limits in the energy range $0.5<\abs{\alpha}^2<3$. Here we find that the coherent state rotation and Legendre states perform best}
\end{figure*}

We plot the performance of the coherent-state-rotation-based receiver in Figs.~\ref{lowenergy} and \ref{highenergy}. In the low-energy regime ($0 < \abs{\alpha}^2 < 0.47$), the coherent state rotation is slightly outperformed by the cat-state-based rotation. At around $\abs{\alpha}^2 = 1.4$, its performance is overtaken by that of the Kennedy receiver. For details on the optimal scaling of the parameter $\beta$ for the cat and coherent state rotations, see App.~\ref{ParamApp}.

\section{Continuously labelled non-Gaussian measurements based on orthogonal polynomials}\label{LegPolSec}
So far, we have considered continuously labelled measurements that naturally extend Gaussian measurements via non-Gaussian unitary preprocessing. However, one may also consider schemes that are not necessarily related to Gaussian measurements through unitary transformations. This motivates the following definition:
\begin{definition}\label{type2}
Type B receivers are defined by the detection of the state which produces a continuous outcome $\psi\in\mathcal{X}$, and is not obtained by unitary preprocessing and Gaussian detection. The POVM elements are given by
\begin{equation}\label{t2povm}
\mathrm{d} \psi\,\hat{\Pi}_{\emph{\textrm{B}}}=\mathrm{d}\psi\,\mu\left(\psi\right) \ket{\psi}\bra{\psi}.
\end{equation}
\end{definition}
Here the measure $\mu(\psi)$ ensures completeness through Eq.~\eqref{ContDiscDiff}. Type B measurements are depicted in Fig.~\ref{type1type2}.

A convenient way to construct such measurements is through families of orthogonal polynomials. For example, the quadrature eigenstates can be expressed in the Fock basis as
\begin{equation}\label{op1}
\ket{x} = \frac{e^{-\frac{x^2}{2}}}{\pi^{1/4}} \sum_{n=0}^\infty \frac{1}{\sqrt{2^n n!}} H_n(x) \ket{n},
\end{equation}
with a completeness relation that follows from the orthogonality of the Hermite polynomials:
\begin{equation}\label{op2}
\frac{1}{\sqrt{\pi}2^n n!} \int_{x \in \mathbb{R}} \mathrm{d}x\, e^{-x^2} H_n(x) H_m(x) = \delta_{nm}.
\end{equation}
This suggests that other classes of orthogonal polynomials satisfying similar orthogonality and completeness conditions may also be used to construct quantum measurements in an analogous manner.

To this end, consider the following Fock basis expansion defined in terms of the Legendre polynomials:
\begin{equation}\label{JacStates}
\ket{s} = \sum_{n=0}^\infty \sqrt{\frac{2n+1}{2}} P_n(s) \ket{n},
\end{equation}
where the Legendre polynomials $P_n(s)$ are given explicitly by
\begin{equation}
P_n(s) = \sum_{m=0}^n \binom{n}{m} \binom{n+m}{m} \left( \frac{s - 1}{2} \right)^m.
\end{equation}
The argument of the Legendre polynomials lies within the interval $s \in [-1,1]$, and owing to their orthogonality, the states defined in Eq.~\eqref{JacStates} satisfy the completeness relation:
\begin{equation}\label{JacPOVM}
\int_{-1}^1 \mathrm{d}s\, \ket{s}\bra{s} = \sum_{n=0}^\infty \ket{n}\bra{n} = \mathbb{I}.
\end{equation}
The three families of classical orthogonal polynomials, Hermite, Laguerre, and Jacobi, are in general not unitarily related (owing to the fact that they are orthogonal on different spaces) \cite{Szego:1975aa}, this implies the lack of existence of a unitary quantum operator which would connect the coefficients of different $\ket{\psi}$ which are given by different families of orthogonal polynomials. Quantum measurements based on Laguerre and Jacobi polynomials are thus not unitarily equivalent to homodyne detection.

We plot the performance of the Legendre states in Fig.~\ref{highenergy}. We find a substantial improvement over homodyne detection, and among all the receivers considered in this paper, the Legendre-state-based receiver yields the second-largest improvement in the high-energy limit. In the energy range $2.2 < \abs{\alpha}^2 < 2.7$, the Legendre polynomial states outperform the coherent state rotations and approach the performance of the Kennedy receiver. Beyond this range, however, their performance deteriorates. For more details on using orthogonal polynomials to construct POVMs, we direct the reader to App.~\ref{PolyApp}.

At first glance, these measurements appear to be purely mathematical constructs. However, there are established results on mode shaping for two-dimensional spatial profiles of light beams \cite{Morizur:2010aa,RevModPhys.92.035005}. In particular, it is known that any input spatial mode can be converted into any desired output mode using reflections and phase plates. How these results extend to our setting is not immediately clear, as we are concerned with the discrimination of two single-mode coherent states. Nevertheless, the techniques developed in this area of the literature may offer a pathway to implementing measurements such as those in Eq.~\eqref{JacPOVM}. Moreover, such measurements may prove useful in a variety of quantum information tasks beyond state discrimination.
\begin{table*}\center
\begin{tabular}{cccc}\hline\hline
  \multicolumn{1}{c}{\small Continuously labelled measurement}&\multicolumn{1}{c}{\small$\hat{\Pi}$}&\multicolumn{1}{c}{\small$\mathfrak{r}^*\left(\mathcal{M}\right)$}&  \multicolumn{1}{c}{\small Near-optimal?}\\
 \hline
\small Homodyne&\small $\mathrm{d}x\, \ket{x}\bra{x}$& \small $ 0$ &\small No\\
\small Heterodyne&\small $ \mathrm{d}^2\beta\, \frac{1}{\pi}\ket{\beta}\bra{\beta}$& \small $0$ &\small No\\
\small Displaced Fock state detection&\small $\mathrm{d}^2\beta\, \frac{1}{\pi}\hat{D}^\dagger (\beta)\ket{n}\bra{n}\hat{D} (\beta)$& \small $ n$ &\small No\\
\small Cubic phase gate + homodyne&\small $\mathrm{d}x\, e^{-\mathrm{i}\gamma \hat{p}^3}\ket{x}\bra{x}e^{\mathrm{i}\gamma \hat{p}^3}$& \small $ \infty$ &\small No\\
\small Kerr + homodyne & \small $\mathrm{d}x\, \hat{U}_\textrm{Kerr}^\dagger \ket{x_{\theta}}\bra{x_\theta} \hat{U}_{\textrm{Kerr}}$ & \small $\infty$& \small Yes\\
\small Fock state rotation + homodyne&\small $\mathrm{d}x\,  \left[\hat{U}^{(N)}_{\left\{\ket{k}\right\}}\left(\vec{\theta}\right)\right]^\dagger\ket{x}\bra{x} \hat{U}^{(N)}_{\left\{\ket{k}\right\}}\left(\vec{\theta}\right)$& \small $ \infty$ &\small Yes\\
\small Cat state rotation + homodyne&\small $\mathrm{d}x\,  \left[\hat{U}^{(N)}_{\left\{\ket{\textrm{cat}(\beta_k,\varphi_k)}\right\}}\left(\vec{\theta}\right)\right]^\dagger\ket{x}\bra{x} \hat{U}^{(N)}_{\left\{\ket{\textrm{cat}(\beta_k,\varphi_k)}\right\}}\left(\vec{\theta}\right)$& \small $ \infty$ &\small Yes\\
\small Coherent state rotation + homodyne&\small $\mathrm{d}x\,  \left[\hat{U}^{(N)}_{\left\{\ket{\beta_k}\right\}}\left(\vec{\theta}\right)\right]^\dagger\ket{x}\bra{x} \hat{U}^{(N)}_{\left\{\ket{\beta_k}\right\}}\left(\vec{\theta}\right)$&\small $ \infty$ &\small Yes\\
\small Laguerre polynomial state detection&\small $\mathrm{d}r\,  \ket{r;\nu}\bra{r;\nu}$&$\geq 1$&\small Yes\\
\small Legendre polynomial state detection&\small $\mathrm{d}s\,  \ket{s}\bra{s}$&\small $\geq 1$&\small Yes\\
\hline\hline
\end{tabular}
\caption{\label{POVMtab}A summary of the continuously labelled measurement schemes studied in this paper, their POVM elements, and the maximum stellar rank \eqref{StelEx} of the binary POVM, $\mathcal{M}=\left\{\hat{\Pi}_1, \hat{\Pi}_2\right\}$. In addition to the near-optimal performance of using a Kerr gate with homodyne detection \cite{PhysRevA.51.1702,Sasaki-Usuda:1995aa}, of the measurements we considered, we found improvement over the Gaussian limit by the generalised states based on the Legendre polynomials \eqref{JacStates} and Laguerre (see App.~\ref{PolyApp}), as well as the unitary rotations followed by homodyne detection, described in \eqref{UFrec1}. Here $\ket{x}, x \in \mathbb{R}$ refers to a quadrature eigenstate, $\ket{\beta}, \beta \in \mathbb{C}$, a coherent state, $\ket{n}, n\in \mathbb{N}$, a Fock state, $\ket{s}$, $s\in[-1,1]$, a Legendre polynomial state, and $\ket{r;\nu}$, $r\in [0,\infty)$, $\nu\in [-1,\infty)$, a Laguerre polynomial state.}
\end{table*}

\section{Non-optimal continuously labelled non-Gaussian measurements}\label{StelRan}
Though we know that non-Gaussianity is required for near-optimal discrimination of two coherent states, we do not in general have a quantification for how much is required, nor a precise characterisation of its form. Moreover, non-Gaussianity is not always helpful in state discrimination. 

To see this we must consider some measure of quantum non-Gaussianity. Several quantifications of non-Gaussianity of quantum states have appeared in the literature. These include: the Wigner function negativity \cite{PhysRevA.98.052350}, the quantum relative entropy \cite{PhysRevA.100.012333}, and robustness measures \cite{PhysRevLett.126.110403}. 

For our purposes, the measure we will use is the recently introduced stellar rank \cite{PhysRevLett.124.063605}, a measure which may be interpreted as the number of photon-additions needed to be applied to a reference Gaussian state. This interpretation provides a natural taxonomy of non-Gaussian states by partitioning the infinite-dimensional Hilbert space of states into classes with the same stellar rank.

For a pure state in an infinite-dimensional Hilbert space the stellar function is defined by
\begin{equation}
 F^*_\psi (\alpha)=e^{\frac{1}{2}\abs{\alpha}^2}\bra{\alpha^*}\ket{\psi}=\sum_{n=0}^\infty\psi_n \frac{\alpha^n}{\sqrt{n!}}, \quad \forall \alpha\in\mathbb{C}.
\end{equation}
The stellar rank, $r^*(\psi)$, is equal to the number of zeroes in the function $F^*_\psi (\alpha)$. For our purposes, we are interested in characterising non-Gaussian measurements, for which we use the convex roof construction for mixed states,
\begin{equation}\label{ConRoof}
\begin{split}
 &\hat{\rho}=\int\mathrm{d}\gamma\, p(\gamma)\ket{\psi(\gamma)}\bra{\psi(\gamma)},\\
 &r^*(\hat{\rho})=\inf_{\left\{p(\gamma), \ket{\psi(\gamma)}\right\}}\sup_\gamma \left\{r^*(\psi(\gamma))\right\},
\end{split}
\end{equation}
and we extend this definition to POVM elements. As we are dealing with binary POVMs, we will take the maximum stellar rank between the two elements. In other words, for a POVM we define the following:

\begin{definition}\label{def1}
The stellar rank of a dichotomic observable, $\mathcal{M}=\left\{\hat{\Pi}_1, \hat{\Pi}_2\right\}$, is given by the maximum of the stellar ranks of the two POVM elements
\begin{equation}\label{StelEx}
\mathfrak{r}^*\left(\mathcal{M}\right)=\max \left\{r^*\left(\hat{\Pi}_1\right), r^*\left(\hat{\Pi}_2\right)\right\}.
\end{equation}
\end{definition}
We recover the consistency condition that Gaussian measurements have $\mathfrak{r}^*\left(\mathcal{M}_{\textrm{G}}\right)=0$, because each element is represented by a mixture of Gaussian states, which satisfy $r^*\left(\hat{\Pi}_1\right)= r^*\left(\hat{\Pi}_2\right)=0$, and the optimal error rate in this scenario is achieved by homodyne detection. For any non-Gaussian improvement in error rates to occur, we must have $\mathfrak{r}^*\left(\mathcal{M}\right)>0$ by Def.~\ref{def1}.

Applying Eq.~\eqref{StelEx} to on-off photon detection, $\mathcal{M}_{\textrm{on/off}}=\left\{\hat{\Pi}_\textrm{off},\hat{\Pi}_\textrm{on}\right\}$, we find that $\mathfrak{r}^*\left(\mathcal{M}_{\textrm{on/off}}\right)=\infty$, due to the fact that  $r^*\left(\hat{\Pi}_\textrm{on}\right)=\infty$. This exposes one of the essential differences between continuously and discretely labelled measurements. For continuously labelled measurements we strictly have $\mathfrak{r}^*\left(\mathcal{M}\right)=r^*\left(\hat{\Pi}_1\right)=r^*\left(\hat{\Pi}_2\right)$ and both elements always have infinite rank (in the functional analysis sense), while for discretely labelled measurements these properties may not be true. More details on the stellar rank can be found in App.~\ref{StelApp}.

Proceeding with definition \eqref{StelEx} we examine the non-Gaussian resources needed for near-optimal coherent state discrimination. All known optimal receivers, have infinite stellar rank because they are based on on-off photon detection. For continuously labelled measurements, the results are summarised in Tab.~\ref{POVMtab}. We find that the receivers based on unitary non-Gaussian rotations all have infinite stellar rank (Eq.~\eqref{UFrec1}), while the receivers based on orthogonal polynomials (Eq.~\eqref{JacPOVM}) have non-zero stellar rank, but we are not able to precisely determine all of the zeroes of their stellar functions (see App.~\ref{PolyApp}). 

We remark that recent work has introduced the $\epsilon$-approximate stellar rank to remedy the problem that many states of interest in continuous-variable systems have a stellar rank $r^*(\psi)=\infty$ \cite{hahn2025assessingnongaussianquantumstate}. We encounter a similar conceptual problem for the measurement schemes we propose in this work, where we find that all optimal and near-optimal measurement schemes imply infinite stellar rank. However, applying the $\epsilon$-approximate stellar rank to measurement operators has not been explored and is beyond the scope of this work.
\subsection{Displaced Fock states}
\begin{figure}\center
    \includegraphics[scale=0.65]{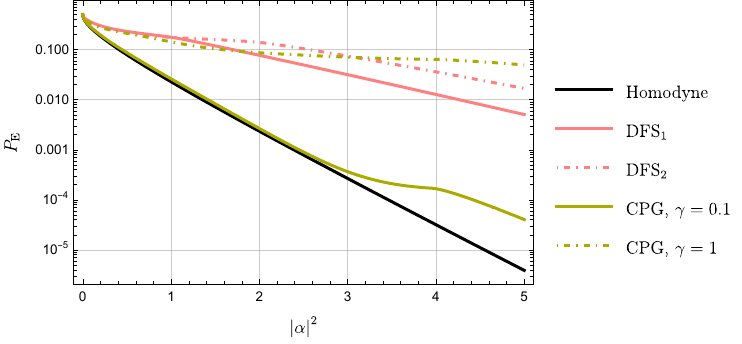} 
    \caption{Performance of the $\textrm{DFS}_n$-based receivers $\hat{\Pi}_{\beta;1}$ and $\hat{\Pi}_{\beta;2}$, and CPG-based receivers for $\gamma=0.1$ and $\gamma=1$ versus homodyne detection.}%
    \label{Pacs}%
\end{figure}

To illustrate that non-Gaussianity is not always useful, consider the displaced Fock states denoted by $\textrm{DFS}_n$, where $n$ indicates the occupation number of the Fock state. $\textrm{DFS}_n$ are defined as follows \cite{Boiteux:1973aa,Wunsche:1991aa}:
\begin{equation}\label{PACSdef}
  \ket{\beta;n}=\hat{D}(\beta)\ket{n},
\end{equation}
and, owing to their completeness relations (see App.~\ref{StelApp}), we may construct the following POVM, $\mathcal{M}_{\textrm{DFS}_n}$, which forms a type B measurement with elements
\begin{equation}
 \mathrm{d}^2\beta\,\hat{\Pi}_{\beta;n}=\mathrm{d}^2\beta\, \frac{1}{\pi}\hat{D}^\dagger (\beta)\ket{n}\bra{n}\hat{D} (\beta).
\end{equation}
For finite $n$, these have finite stellar rank $\mathfrak{r}^*\left(\mathcal{M}_{\textrm{DFS}_n}\right)=n$. They also closely resemble the Kennedy receiver, the only difference being that we resolve the identity over the phase-space variable $\beta$, and not over the occupation number $n'$. Explicitly, this is:
\begin{equation}
\int_{\mathbb{C}}\mathrm{d}^2\beta\,\hat{\Pi}_{\beta;n}=\sum_{n'}\hat{D}^\dagger (\alpha)\ket{n'}\bra{n'}\hat{D} (\alpha)=\mathbb{I}.
\end{equation}
Despite this, measurements with $\textrm{DFS}_n$ perform worse than the Gaussian limit. Moreover, for sufficiently large $\abs{\alpha}^2$, the performance degrades with increasing $n$ for all values of $n$ that we tested. This behaviour is shown in Fig.~\ref{Pacs} where we compare homodyne detection and $\textrm{DFS}_n$ with $n=1,2$. We remark that the $\textrm{DFS}_n$ resemble the $n$-photon-added coherent states \cite{PhysRevA.43.492,PhysRevA.46.485}, the difference being the ordering of the displacement and the photon additions in their definition. Thus, the $n$-photon-added coherent states do not define a suitable measurement basis as they do not form an overcomplete set of states.

We conjecture that the form of the probability distributions for all finite stellar rank measurements is $\propto \abs{R_{n}\left(\beta\pm\alpha\right)}^2 G_{\pm\alpha,\sigma}\left(\beta\right)$, where $R_n(x)$ is a polynomial of degree $n$ in $x$, and $G_{\mu,\sigma}\left(x\right)$ is a Gaussian distribution of mean $\mu$ and variance $\sigma$ in $x$. It is unclear whether modulating Gaussians by such polynomials can increase their total variation distance, and consequently, whether error rates lower than the Gaussian limit can be achieved by Eq.~\eqref{totvardist}.

\subsection{Cubic phase gate}
Infinite stellar rank also does not ensure near-optimal performance in coherent state discrimination. To see this, consider the cubic phase gate (CPG). The CPG is the canonical non-Gaussian element in optical quantum computing \cite{PhysRevResearch.6.023332}. It represents the lowest-order non-Gaussian element (or gate in computing nomenclature) which when added to the Gaussian gate set can enable universal optical quantum computing. Despite the difficulties in its experimental realisation \cite{Sakaguchi:2023aa,PhysRevA.93.022301,PhysRevResearch.3.043026}, the utility of the CPG remains an active area of research both theoretically and practically.

In the present context, we are interested in its utility in coherent state discrimination. The CPG is given by
\begin{equation}
 \hat{U}(\gamma)=e^{\mathrm{i}\gamma \hat{p}^3}, \quad \gamma \in \mathbb{R},
\end{equation}
where $\hat{p}$ is the momentum operator. Applying this to a coherent state and taking the position space wavefunction we find
\begin{equation}\label{cubmain}
\begin{split}
& \bra{x}e^{\mathrm{i}\gamma \hat{p}^3}\ket{\alpha}=\\
&\frac{\left(4\pi\right)^{\frac{1}{4}}}{\abs{3\gamma}^{\frac{1}{3}}}e^{\frac{1}{108\gamma^2}}e^{\frac{x-\sqrt{2}\alpha}{6\gamma}}\textrm{Ai}\left((3\gamma)^{-\frac{1}{3}}\left(x-\sqrt{2}\alpha+\frac{1}{12\gamma}\right)\right).
\end{split}
\end{equation}
Here $\textrm{Ai}(x)$ is the Airy function of the first kind \cite{mabramowitz64:handbook}. Note that typically the CPG is performed on the rotated quadrature, $\hat{x}$. In our case, however, if we were to apply the CPG along $\hat{x}$ and then perform homodyne detection along the same axis, this would amount to an overall phase in the wavefunction and thus would be equivalent to homodyne detection. For details of the derivation, we refer the reader to App.~\ref{cubphasapp}.

A single CPG followed by homodyne detection constitutes a type A measurement, denoted by $\mathcal{M}_\textrm{CPG}$. Computing the stellar rank of the associated measurement operators,
\begin{equation}\label{CubicPOVM}
\mathrm{d}x\, \hat{\Pi}_\gamma= \mathrm{d}x\,  e^{-\mathrm{i}\gamma \hat{p}^3}\ket{x}\bra{x}e^{\mathrm{i}\gamma \hat{p}^3},
\end{equation}
yields $\mathfrak{r}^* \left(\mathcal{M}_\textrm{CPG}\right)=\infty$. This can be inferred from the presence of an Airy function, which contains infinitely many zeroes, in the stellar function of \eqref{CubicPOVM}. We plot the performance of the CPG in Fig.~\ref{Pacs}. We find that for increasing cubicity, $\gamma$, the error rate degrades. Thus, infinite stellar rank is also not enough to ensure improved performance in coherent state discrimination, beyond the Gaussian limit.

We emphasise that a single application of the cubic phase gate is ineffective. Moreover, the addition of further Gaussian unitaries to Eq.~\eqref{CubicPOVM} did not numerically improve discrimination beyond the Gaussian limit, as suggested by the results on Gaussian operations combined with CPGs and photon detection in Ref.~\cite{warke2024photonicquantumreceiverattaining}. We conjecture that a single CPG is not useful for coherent state discrimination when combined with gaussian operations. At first glance, this may seem to contradict results on decomposing arbitrary gates into products of CPGs and Gaussian gates \cite{Kalajdzievski2021exactapproximate}, however, the decomposition of an arbitrary non-Gaussian gate, such as those in Eq.~\eqref{NGrotU}, requires many CPGs. Consequently, this does not imply that we can approximate a useful non-Gaussian gate for continuously labelled state discrimination using only a single CPG. Further discussion and supporting numerical evidence is contained in App.~\ref{cubphasapp}.

The stellar rank thus gives us a primitive way to characterise the non-Gaussianity needed for near-optimal coherent state discrimination. Under certain circumstances it can provide some criteria for distinguishing some continuously labelled measurement types from e.g. on-off photon detection. It does not, however, provide a complete quantification. Although currently lacking, a non-Gaussian resource monotone which is able to do this, as well as inheriting a structure like the stellar hierarchy, would be important for studying non-Gaussianity in more general quantum information contexts.

\section{Conclusion}\label{conclusion}
We have explored continuously labelled non-Gaussian operations that enable near-optimal discrimination of two coherent states. We have considered two distinct types of receivers which are unitarily inequivalent: type A based on non-Gaussian unitaries combined with homodyne detection, and type B based on properties of orthogonal polynomials.

Type A receivers were found to be near-optimal for large ranges of the signal state energy $\abs{\alpha}^2$, for three distinct subclasses of receiver, based on rotations by cat states, coherent states, and Fock states. Fock state rotations, known as SNAP gates, have been realised experimentally. For one of our proposed schemes, namely the a coherent state rotation, we provided a decomposition into a single SNAP gate and two displacements. Given that SNAP gates are already realisable with current technology \cite{PhysRevLett.115.137002}, we anticipate that a coherent state rotation combined with a homodyne detector could also be implemented using existing methods. It would therefore be worthwhile to assess its practical viability, as such a scheme performs as well as, and in some cases better than, other near-optimal photon detection-based approaches. Type B receivers were also found to be near-optimal for larger ranges of the signal state energy, for two distinct subclasses, based on the Legendre and Laguerre polynomials. Moreover, all near-optimal receivers we constructed were able to beat the Kennedy receiver limit for significant energy ranges.

We have also provided a quantification of the non-Gaussianity of the measurements, based on the stellar rank, which provides a hierarchy of non-Gaussian quantum measurements. Specifically, Gaussian measurements have stellar rank $\mathfrak{r}^*\left(\mathcal{M}_\textrm{G}\right)=0$ and all other measurements have $\mathfrak{r}^*\left(\mathcal{M}_\textrm{G}\right)>0$. All known optimal and near-optimal solutions, including the ones presented in this work (up to proofs about the zeroes of the stellar functions of the type B receivers), require $\mathfrak{r}^*\left(\mathcal{M}\right)=\infty$. 

Our schemes, which involve a finite number of rotations by specific classes of states, were unable to saturate the Helstrom bound. Whether a purely continuously labelled measurement can achieve this remains an open question. One possible approach is to approximate a known optimal solution, which involves photon detection, as closely as possible using continuously labelled measurements. Should such a measurement exist, techniques for arbitrary gate synthesis in continuous-variable systems may enable the implementation of the required operations, thereby paving the way towards a practical realisation \cite{Arrazola_2019, Kalajdzievski2021exactapproximate}.

The scheme based on the Legendre polynomials performed commensurably with the receivers based on non-Gaussian rotations and homodyne detection in the high-energy limit. We considered only some of the classical orthogonal polynomials but there is a vast literature on generalised orthogonal polynomials such as those organised by the Askey scheme and their $q-$analogues \cite{koekoek1996askeyscheme}, some of which may have mathematical utility in quantum information tasks, like state discrimination. If they turn out to be of theoretical importance, perhaps further study on their physical realisation may be of interest. 

It would be interesting to investigate whether infinite stellar rank is necessary for achieving near-optimal and optimal discrimination of two coherent states. This could provide a pathway to a more complete resource theory of non-Gaussian quantum measurements. A fully developed resource theory may necessitate a distinction between continuously and discretely labelled measurements, from which one could identify tasks which may be possible with one class and not the other. Recent work addresses infinite stellar rank states by approximating them with lower stellar rank states up to some precision quantified by the stellar fidelity \cite{hahn2025assessingnongaussianquantumstate}. These results are not immediately applicable to our case, but there may exist suitable modifications to allow for a finer-grained description of the non-Gaussianity required in state discrimination. These questions will be explored in future work.

The schemes we developed are near-optimal for coherent states. For the discrimination of two arbitrary Gaussian states, an optimal scheme (outside of the projective Helstrom measurement) is still unknown \cite{Olivare,5407627}. It may be useful to compare our scheme for discriminating more general Gaussian states with other currently known near-optimal schemes \cite{Notarnicola:23,PhysRevLett.117.200501}. Moreover, since heterodyne detection has been shown to be optimal among Gaussian measurements for quadrature phase-shift keying \cite{60560,Muller:2012aa}, it would be worthwhile to revisit the applicability of the receivers proposed in this work within that setting.

\begin{acknowledgments}
J.M., S.K., and H.K. are supported by KIAS individual grant numbers QP088702 (J.M.), CG086202 (S.K.), and CG085302 (H.K.) at the Korea Institute for Advanced Study. This work was supported by the Institute of Information \& Communications Technology Planning \& Evaluation (IITP) grant, funded by the Korea government (MSIT) (No. 2022-0-00463). 
\end{acknowledgments}

\appendix

\section{Inequivalence of photon detection and homodyne and heterodyne detection}\label{HomIneq}
Here we prove the inequivalence of photon detection and homodyne and heterodyne detection. We will show the following:
\begin{equation}\label{hom}
\begin{split}
 &\int_{\psi \in\mathcal{R}}\mathrm{d} \psi\, \mu(\psi) \ket{\psi}\bra{\psi} \not = \ket{0}\bra{0}\\
 &=\sum_{n=0}^\infty \sum_{m=0}^\infty \delta_{nm}\delta_{n0}\ket{n}\bra{m}, \quad \forall \mathcal{R},
 \end{split}
\end{equation}
for the instances where $\ket{\psi}=\ket{x}, x\in \mathbb{R}, \mu(\psi)=1$ is a quadrature eigenstate, and $\ket{\psi}=\ket{\beta}, \beta \in \mathbb{C}, \mu(\psi)=\frac{1}{\pi}$, is a coherent state. The left-hand side of \eqref{hom} may be expanded in diagonal and off-diagonal parts as
\begin{equation}
\begin{split}
 &\int_{\psi \in\mathcal{R}}\mathrm{d} \psi\,  \mu(\psi)\ket{\psi}\bra{\psi}\\
 &=\int_{\psi \in\mathcal{R}}\mathrm{d} \psi\,  \mu(\psi) \sum_{n=0}^\infty \sum_{m=0}^\infty \bra{n}\ket{\psi}\bra{\psi}\ket{m}\ket{n}\bra{m}\\
 &=\int_{\psi \in\mathcal{R}}\mathrm{d} \psi\,  \mu(\psi)\sum_{n=0}^\infty  \abs{\bra{n}\ket{\psi}}^2 \ket{n}\bra{n}+\textrm{off-diagonal}\\
 &= \sum_{n=0}^\infty \mathcal{I}_{nn}+\textrm{off-diagonal},
 \end{split}
\end{equation}
where
\begin{equation}\label{diagel}
  \mathcal{I}_{nn}= \int_{\psi \in\mathcal{R}}\mathrm{d} \psi\,  \mu(\psi)\abs{\bra{n}\ket{\psi}}^2 \ket{n}\bra{n}.
\end{equation}
\subsection{Homodyne detection}
First, consider $\ket{\psi}=\ket{x}$, a quadrature eigenstate. For any $\mathcal{R}\subseteq \mathbb{R}$, its diagonal elements \eqref{diagel} are given by
\begin{equation}\label{integrand2}
 \mathcal{I}_{nn}= \left(\frac{1}{\pi}\right)^{\frac{1}{2}}\int_{x\in\mathcal{R}}\mathrm{d}x\, \frac{1}{2^n n!}e^{-x^2}H_n^2(x)\ket{n}\bra{n}.
\end{equation}

Here we see that the integrand of \eqref{integrand2} is an even function and is positive $\forall x$, meaning we can upper-bound by a larger integration region $\tilde{\mathcal{R}}$ containing $\mathcal{R}$, such that
\begin{equation}\label{integrand3}
\tilde{\mathcal{I}}_{nn}= \left(\frac{1}{\pi}\right)^{\frac{1}{2}}\int_{x\in\tilde{\mathcal{R}}}\mathrm{d}x\, \frac{1}{2^n n!}e^{-x^2}H_n^2(x)\ket{n}\bra{n},
\end{equation}
and
\begin{equation}\label{homineq}
 \bra{n}\mathcal{I}_{nn}\ket{n} \leq  \bra{n}\tilde{\mathcal{I}}_{nn}\ket{n}\leq \abs{\bra{n}\ket{n}}^2=1,
\end{equation}
corresponding to the integration regions $\mathcal{R}\subseteq \tilde{\mathcal{R}}\subseteq \mathbb{R}$, where the right-hand side of \eqref{homineq} comes from the orthogonality relations of the Hermite polynomials, i.e., the integration region $x\in \mathbb{R}$. We see that in order to get a unit projection onto the ground state $\bra{0}\mathcal{I}_{00}\ket{0}$ we need to take the set $\mathcal{R}=\mathbb{R}$, i.e., the resolution of the identity for the quadrature eigenstates. Thus we conclude that vacuum state projection is not possible by any splitting of integration regions in homodyne detection.
\subsection{Heterodyne detection}
Next we consider $\ket{\psi}=\ket{\beta}$. The diagonal elements \eqref{diagel} in radial coordinates are
\begin{equation}\label{integrand1}
 \mathcal{I}_{nn}= 2\int_{x\in\mathcal{R}}\mathrm{d}r\, \frac{r^{1+2n} e^{-r^2}}{n!}\ket{n}\bra{n}.
\end{equation}
Given that there is no explicit angular dependence in the integral and the integrand is positive on $r\in [0,\infty)$, we may enclose the region $\mathcal{R}$ by a larger region $\tilde{\mathcal{R}}$ containing $\mathcal{R}$ and the same chain of inequalities as \eqref{homineq} follows with unit projection onto the vacuum state this time only being possible with $\mathcal{R}=\mathbb{C}$, which just recovers the completeness relation for the coherent states. Thus no splitting of integration regions in heterodyne detection can reproduce on-off photon detection.
\section{Rotating by Fock states with increasing $N$}\label{IncN}
Consider the unitary operator Eq.~\eqref{NGrotU} with rotations by Fock states, i.e., SNAP gates \cite{PhysRevLett.115.137002, Park_2024},
\begin{equation}
 \left\{\ket{\psi_k}\right\}=\left\{\ket{k}\right\}.
\end{equation}
To obtain some concrete results we perform numerical optimisations for various values of $N$, demonstrating that increasing $N$ can improve the error rate of receivers based on unitary rotations by pure state projectors, as described in Sec.~\ref{UnitSec}. Though these receivers allow us to enter the near-optimal regime, they are ultimately outperformed by other continuously labelled receivers considered in Tab.~\ref{POVMtab}.

\begin{figure}
    \centering
    \includegraphics[scale=0.65]{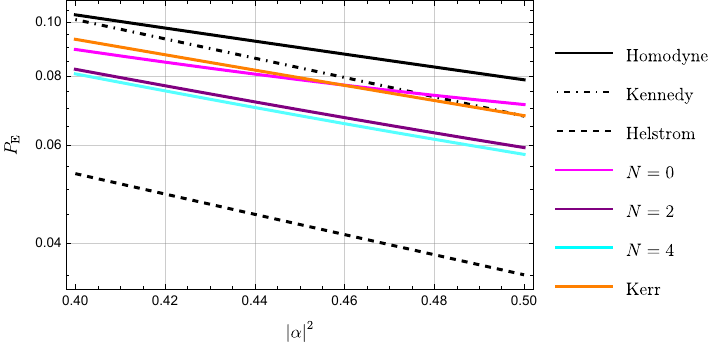} 
    \caption{Error rates for Fock state rotation based receivers, $\hat{\Pi}^{(N)}_{\left\{\ket{k}\right\}}$, for different values of $N$ with $\abs{\mathcal{S}_1}=1, \abs{\mathcal{S}_2}=3$, and $\abs{\mathcal{S}_3}=5$ corresponding to the sets in \eqref{focksets}. We find that the error rate decreases with increasing $N$. The performance of the receiver in \cite{PhysRevA.51.1702} based on a Kerr gate combined with homodyne detection is also shown.}%
    \label{FockN}%
\end{figure}

The performance of a receiver with $N$ rotations can, at worst, be equal to that of a receiver with $N-1$ rotations, since we may set the $N^{\textrm{th}}$ rotation parameter to $\theta_N=0$. Here we show an explicit example. Consider the following three cases:
\begin{equation}\label{focksets} 
\begin{split}
 \mathcal{S}_1&=\left\{ \ket{1}\right\},\\
  \mathcal{S}_2&=\left\{\ket{0},\ket{1}, \ket{2} \right\},\\
   \mathcal{S}_3&=\left\{\ket{0},\ket{1},\ket{2},\ket{3},\ket{4}\right\},
\end{split}
\end{equation}
where the optimal rotation angles, $\vec{\theta}_{\mathcal{S}_i} (\alpha)$, are determined numerically. Over the range of $\abs{\alpha}^2$ considered, we observe that $\vec{\theta}_{\mathcal{S}_1} (\alpha)=(\pi)$, $\vec{\theta}_{\mathcal{S}_2} (\alpha)=(\pi,\pi,\pi)$, and $\vec{\theta}_{\mathcal{S}_3} (\alpha)$ is variable. Their respective error rates are plotted in Fig.~\ref{FockN}. The error rates decrease with increasing $N$ and as such we conclude that increasing $N$ can improve performance. We also find that the receiver corresponding to $N=0$ performs comparably with the Kerr gate and homodyne detection based-receiver considered in \cite{PhysRevA.51.1702,Sasaki-Usuda:1995aa}.

\section{Optimal parameters for coherent state and cat state rotations}\label{ParamApp}
\begin{figure}\center
    \includegraphics[scale=0.65]{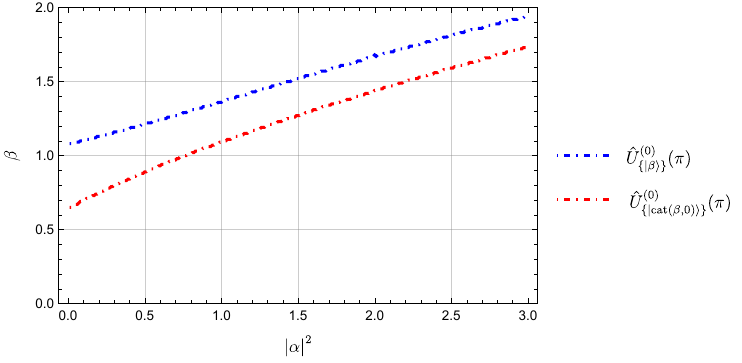} 
    \caption{The optimal values of $\beta$ for both cat and coherent state rotations, $ \hat{U}^{(0)}_{\left\{\ket{\beta}\right\}}\left(\pi\right)$ and $\hat{U}^{(0)}_{\left\{\ket{\textrm{cat}(\beta)}\right\}}\left(\pi\right)$ in Eq.~\eqref{NGrotU}, as a function of the input state intensity, $\abs{\alpha}^2$, for $0.01\leq\abs{\alpha}^2\leq 3$.}%
    \label{ScalGraph}%
\end{figure}

Here we show that the cat and coherent state parameters, $\beta$, follow an approximately linear scaling in the coherent state intensity. In Fig.~\ref{ScalGraph} we plot the optimal values of $\beta$ for both the coherent state and cat state rotations as a function of $\abs{\alpha}^2$, with a spacing between points of $\Delta \abs{\alpha}^2=0.01$. Over the energy ranges considered in the paper, $0<\abs{\alpha}\leq 3$, we found that a linear fit on the coherent states has a scaling of $1.07+0.296\abs{\alpha}^2$ and the cat states have a fit of $0.715+0.355\abs{\alpha}^2$. We remark that the linear fit is more accurate for the coherent state rotations. Note that we exclude the point $\abs{\alpha}^2=0$ because this corresponds to discriminating two identical vacuum states, for which no unitary preprocessing can help.
\section{Orthogonal polynomial states}\label{PolyApp}
\begin{figure}\center
    \includegraphics[scale=0.65]{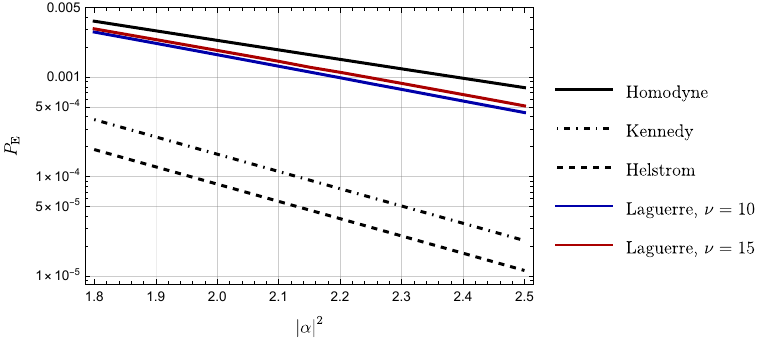} 
    \caption{Performance of the Laguerre polynomial, $L_n^{(\nu)}(x)$, receivers defined by Eqs.~\eqref{PolyStat} and \eqref{polycomplete}, for $\nu=10$ and $\nu=15$. We find a marginal improvement over the Gaussian limit indicating a non-Gaussian advantage.}%
    \label{LagGraph}%
\end{figure}
\begin{table*}\center
\begin{tabular}{ccc}\hline\hline
\small Polynomial &\small$P_n(x)$ &  \small$w_n(x)$\\
\hline
\small Hermite $H_n(x)$& \small$(-1)^n e^{x^2}\frac{\mathrm{d}^n}{\mathrm{d}x^n} e^{-x^2}$ & \small$ \frac{e^{-x^2}}{\sqrt{\pi}2^n n!}$\\
\small Laguerre $L_n^{(\nu)}(x)$&\small$\frac{x^{-\nu}e^x}{n!}\frac{\mathrm{d^n}}{\mathrm{d}x^n} \left(e^{-x}x^{n+\nu}\right)$ &\small $\frac{n! x^\nu e^{-x}}{\Gamma(n+\nu+1)}$\\
\small Jacobi $J_n^{(\nu,\mu)}(x)$&\small$\frac{(-1)^n(1-x)^{-\nu}(1+x)^{-\mu}}{2^n n!}\frac{\mathrm{d}^n}{\mathrm{d}x^n}\left( (1-x)^\nu (1+x)^\mu (1-x^2)^n\right)$ & \small$\frac{(1-x)^\nu(1+x)^\mu (2n+\nu+\mu+1)\Gamma(n+\nu+\mu+1)n!}{2^{\nu+\mu+1}\Gamma(n+\nu+1)\Gamma(n+\mu+1)}$\\
\hline\hline
\end{tabular}
\caption{\label{PolyTab}Classical orthogonal polynomials, their Rodrigues' formulas, and their weight functions. The generalised Laguerre and Jacobi polynomials come with additional parameters satisfying $\nu,\mu>-1$. Their orthogonality relations are given by Eq.~\eqref{PolyOrthog}.}
\end{table*}
Classical orthogonal polynomials, $P_n(\psi)$, satisfy the following orthogonality conditions
\begin{equation}\label{PolyOrthog}
 \int_{\psi \in \mathcal{X}}\mathrm{d}\psi\, w_n(\psi) P_n(\psi)P_m(\psi)=\delta_{nm},
\end{equation}
where $w_n(\psi)$ is some positive weight function. There are three distinct families of classical orthogonal polynomials, these are: the Hermite polynomials, orthogonal on $\mathcal{X}=\mathbb{R}$; the Laguerre polynomials, orthogonal on $\mathcal{X}=[0,\infty)$; and the Jacobi polynomials, orthogonal on $\mathcal{X}=[-1,1]$ \cite{koekoek1996askeyscheme,Szego:1975aa}. The essential properties of the three families of classical orthogonal polynomials are summarised in Tab.~\ref{PolyTab}.

Property \eqref{PolyOrthog} allows us to construct generalised quantum states,
\begin{equation}\label{PolyStat}
 \ket{\psi}=\sum_{n=0}^\infty \sqrt{w_n(\psi)}P_n(\psi)\ket{n},
\end{equation}
which satisfy the completeness relation
\begin{equation}\label{polycomplete}
 \int_{\psi \in \mathcal{X}}\mathrm{d}\psi\, \ket{\psi}\bra{\psi}=\sum_{n=0}^\infty \ket{n}\bra{n}=\mathbb{I}.
\end{equation}
When we choose the Hermite polynomials, the generalised quantum states \eqref{PolyStat} become the quadrature eigenstates and these also define the Gaussian limit for coherent state discrimination. If the generalised Laguerre polynomials are chosen, the generalised quantum states resemble Laguerre-Gauss modes \cite{Simpson:97,PhysRevLett.88.053601}, up to a rescaling of the radial variable. We found a very slight non-Gaussian advantage using these states as reported in Tab.~\ref{POVMtab}. When the Jacobi polynomial states are chosen, these do not appear in the literature, perhaps because a variable constrained to the interval $[-1,1]$ lacks an obvious physical interpretation. Nevertheless, we can define measurements with these states, and when the parameters $\nu=\mu=0$, the Jacobi polynomials are precisely the Legendre polynomials studied in the main text in Sec.~\ref{LegPolSec}.

Given the improvement beyond the Gaussian limit, we know that the generalised polynomial states have non-Gaussian character, though determining their stellar rank involves counting the zeroes of the function
\begin{equation}
 F^*_{P}(\alpha)=\sum_{n=0}^\infty\frac{\sqrt{w_n(\psi)}P_n(\psi)}{\sqrt{n!}}\alpha^n,
\end{equation}
which for the Laguerre and Jacobi polynomials we are not able to do. We know, however, given the observed non-Gaussian improvement, these functions must have at least one zero.

In Fig.~\ref{LagGraph} we plot the error rate for a receiver based on generalised Laguerre polynomials for the parameters $\nu=10$ and $\nu=15$. We observe a slight improvement over the homodyne limit, verifying their non-Gaussian character, but the improvement is only marginal.

\section{Stellar rank and displaced Fock states}\label{StelApp}
Here we briefly introduce the stellar rank and the stellar hierarchy of non-Gaussian states. Described in \cite{PhysRevLett.124.063605}, the stellar function for a pure state in an infinite-dimensional Hilbert space is defined by
\begin{equation}
 F^*_\psi (\alpha)=e^{\frac{1}{2}\abs{\alpha}^2}\bra{\alpha^*}\ket{\psi}=\sum_{n\geq0}\psi_n \frac{\alpha^n}{\sqrt{n!}}, \quad \forall \alpha\in\mathbb{C},
\end{equation}
which is closely related to the Husimi Q-distribution \cite{1940264,PhysRev.177.1882} through
\begin{equation}
 Q_\psi(\alpha)=\frac{1}{\pi}\abs{\bra{\alpha}\ket{\psi}}^2=\frac{e^{-\abs{\alpha}^2}}{\pi}\abs{F^*_\psi (\alpha)}^2.
\end{equation}
The stellar rank, $r^*(\psi)$, thus organises non-Gaussian states by partitioning the infinite-dimensional Hilbert space of states into classes with the same number of zeroes in their stellar functions. Any pure state may then be expressed as \cite{PhysRevLett.124.063605}
\begin{equation}
 \ket{\psi}= F^*_\psi (\hat{a}^\dagger)\ket{0}.
\end{equation}
For a state of finite stellar rank we have the following 
\begin{equation}
  \ket{\psi}=\frac{1}{\mathcal{N}}\left[\prod_{n=1}^{r^*(\psi)}\hat{D}(\beta_n)\hat{a}^\dagger \hat{D}^\dagger (\beta_n)\right]\ket{G_\psi}.
\end{equation}
Here $\ket{G_\psi}$ is a reference Gaussian state and $\mathcal{N}$ is a normalisation. This decomposition is unique up to a reordering of the roots. We see that the stellar rank essentially counts the number of photon-additions needed to create the state $\ket{\psi}$ from a reference Gaussian state. It thus provides a hierarchy of non-Gaussian states. 

For $\textrm{DFS}_n$ we can immediately determine their stellar rank by rewriting Eq.~\eqref{PACSdef} as
\begin{equation}
 \ket{\beta;n}=\frac{1}{\sqrt{n!}}\prod_{i=1}^n\left[ \hat{D}(\beta)\hat{a}^\dagger \hat{D}^\dagger(\beta)\right]\ket{\beta},
\end{equation}
from which it follows that $r^*\left(\hat{\Pi}_{\beta;n}\right)=n$, and hence $\mathfrak{r}^*\left(\mathcal{M}_{\textrm{DFS}_n}\right)=n$. The corresponding completeness relation is given by \cite{Boiteux:1973aa,Wunsche:1991aa}
\begin{equation}\label{pacsroi}
 \int_{\beta \in\mathbb{C}}\frac{\mathrm{d}^2\beta}{\pi}\, \ket{\beta;n}\bra{\beta;n}=\mathbb{I},
\end{equation}
which is valid $\forall n$.

\section{Cubic phase gate}\label{cubphasapp}
\begin{figure*}\center
    \begin{subfigure}[t]{0.3\textwidth}
        \includegraphics[width=\textwidth]{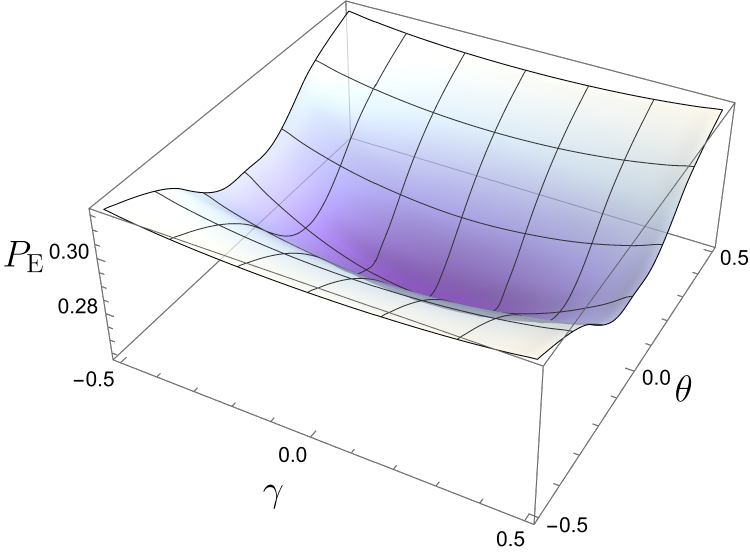}
        \caption{$\abs{\alpha}^2=0.1$}
        \label{fig:sub1}
    \end{subfigure}
    \hfill
    \begin{subfigure}[t]{0.3\textwidth}
        \includegraphics[width=\textwidth]{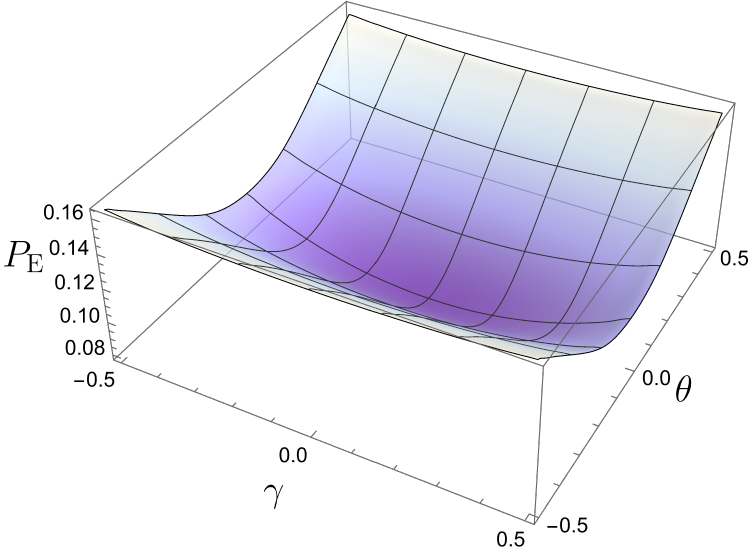}
        \caption{$\abs{\alpha}^2=0.5$}
        \label{fig:sub2}
    \end{subfigure}
    \hfill
    \begin{subfigure}[t]{0.3\textwidth}
        \includegraphics[width=\textwidth]{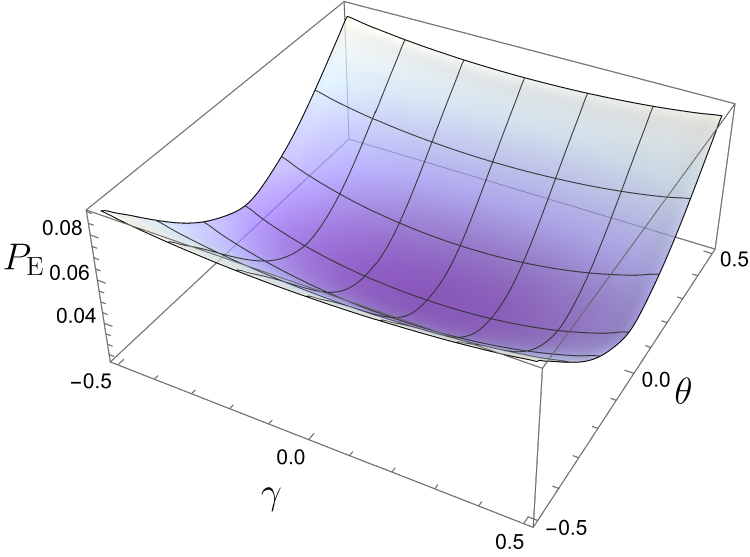}
        \caption{$\abs{\alpha}^2=1$}
        \label{fig:sub3}
    \end{subfigure}
    \caption{The error rates, $P_\textrm{E}$, of a single cubic phase gate combined with homodyne detection plotted against the cubicity, $\gamma$, and the rotation angle, $\theta$, for different values of the coherent state amplitude $\abs{\alpha}^2$. In all cases tested, the minimum error rate is located at the point $\gamma=\theta=0$.}
    \label{CubAll}
\end{figure*}

In order to determine the wavefunction for the CPG, consider the following
\begin{equation}\label{CubWFAPP}
\bra{x;\theta}e^{\mathrm{i}\gamma \hat{p}^3}\ket{\alpha}=\bra{x;\theta}\int_{p\in\mathbb{R}}\mathrm{d}p\,e^{\mathrm{i}\gamma p^3}\ket{p}\bra{p}\ket{\alpha}
\end{equation}
where $\ket{p}$ is a momentum eigenstate and $\ket{x;\theta}= e^{\mathrm{i}\theta \hat{a}^\dagger \hat{a}}\ket{x}$ is a rotated quadrature eigenstate. Using the following property of the fractional Fourier transform \cite{PhysRevA.60.3467}
\begin{equation}
 \bra{x;\theta}\ket{p}=\left(\frac{1+\mathrm{i} \tan \theta}{2\pi}\right)^\frac{1}{2} e^{\mathrm{i}\frac{px}{\cos\theta}}e^{-\frac{\mathrm{i}}{2}(x^2 + p^2)\tan\theta},
\end{equation}
and results contained in \cite{Katori:2009aa}, in particular, the following identity:
\begin{equation}\label{AiId}
\begin{split}
 &\frac{1}{2\pi}\int_{y\in \mathbb{R}} \mathrm{d}y\, \exp\left(\mathrm{i}\frac{x}{c} y -\frac{1}{4c^2}y^2 +\frac{\mathrm{i}}{3} y^3 \right)\\
 &=\exp\left(\frac{1}{4 c^3}\left(x+\frac{1}{24c^3}\right)\right) \mathrm{Ai}\left(\frac{x}{c}+\frac{1}{16c^4}\right),
\end{split}
\end{equation}
which is valid $\forall c \in \mathbb{C}$. We find that the wavefunction \eqref{CubWFAPP} is equal to
\begin{equation}\label{rotcub}
\begin{split}
& \bra{x;\theta}e^{\mathrm{i}\gamma \hat{p}^3}\ket{\alpha}=\\
&\frac{\textrm{Ai}\left((3\gamma)^{-\frac{1}{3}}\left(\frac{x}{\cos\theta}-\sqrt{2}\alpha+\frac{(1+\mathrm{i}\tan\theta)^2}{12\gamma}\right)\right)}{N(\gamma,x,\theta)},
\end{split}
\end{equation}
where
\begin{equation}
\begin{split}
 \frac{1}{N(\gamma,x,\theta)}&=\frac{(4\pi)^\frac{1}{4}}{\abs{3\gamma}^\frac{1}{3}}(1+\mathrm{i}\tan \theta)^\frac{1}{2}e^{-\frac{\mathrm{i}}{2}x^2 \tan \theta}e^{\frac{(1+\mathrm{i}\tan \theta)^3}{108\gamma^2}}\\
 &\times e^{\frac{1}{6\gamma}(1+\mathrm{i}\tan \theta)\left(\frac{x}{\cos\theta}-\sqrt{2}\alpha\right)}.
 \end{split}
\end{equation}
The $\theta \rightarrow0$ limit of Eq.~\eqref{rotcub} recovers the wavefunction quoted in the main text, Eq.~\eqref{cubmain}, i.e. the position wavefunction. From this we define the type A POVM elements
\begin{equation}
\mathrm{d}x\, \hat{\Pi}_{\gamma,\theta}= \mathrm{d}x\, e^{-\mathrm{i}\gamma \hat{p}^3}\ket{x;\theta}\bra{x;\theta}e^{\mathrm{i}\gamma \hat{p}^3},
\end{equation}
which are a generalisation of Eq.~\eqref{CubicPOVM}.

We immediately see the ineffectiveness of $x$ displacements, as introducing an additional displacement of $\delta$ before the CPG amounts to a translation of the centre point between $\ket{\alpha}$ and $\ket{-\alpha}$ to $\ket{\alpha+\delta}$ and $\ket{-\alpha + \delta}$, which leaves Eq.~\eqref{totvardist} invariant. Additionally, for the squeezing operator, $\hat{S}(r)=e^{\frac{r}{2}\left(\left(\hat{a}^\dagger \right)^2 -\hat{a}^2\right)}$ with $r\in \mathbb{R}$, we have the braiding relation $\hat{\Gamma}(\gamma) \hat{S}(r)=\hat{S}(r)\hat{\Gamma}(\gamma e^{-3r})$ \cite{PhysRevA.100.013831}, implying that introducing squeezing is equivalent to a rescaling of the cubicity, and we found no evidence that any cubicity can improve the discrimination probability beyond the Gaussian limit. Lastly, we considered rotations, as encoded in Eq.~\eqref{CubWFAPP}, and found that $\theta=0$ is optimal for improving discrimination error rates. This can be understood from the fact that homodyne detection along the $p-$axis cannot outperform random guessing (since we are considering two coherent states oriented on the $x-$axis), which is the worst possible outcome. Furthermore, homodyne detection along intermediate angles tends to worsen discrimination probabilities. Numerical evidence of this is plotted in Fig.~\ref{CubAll}. We observe that homodyning along any quadrature other than the $x-$axis degrades performance, and performance worsens with increasing rotation angle, $\theta$, and cubicity, $\gamma$.

\section{Comparison with photon detection-based near-optimal solutions}\label{compapp}
In this appendix we show that continuously labelled near-optimal solutions to coherent state discrimination may have a quantitative advantage over photon detection-based near-optimal solutions. We consider the two near-optimal solutions presented in \cite{PhysRevA.78.022320}, called the `type-I' and `type-II' receivers, which both use photon detection and an optimised displacement, and the type-II receiver uses an additional optimised squeezing.
\begin{figure}\center
    \includegraphics[scale=0.65]{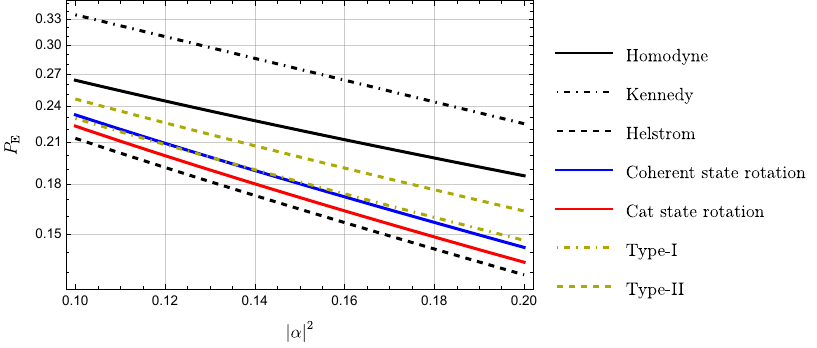} 
    \caption{Performance of the coherent state and cat state rotation-based receivers compared with the type-I and type-II near-optimal photon detection-based receivers in \cite{PhysRevA.78.022320} over the energy range $0.1<\abs{\alpha}^2<0.2$.}%
    \label{no1}%
\end{figure}
\begin{figure}\center
    \includegraphics[scale=0.65]{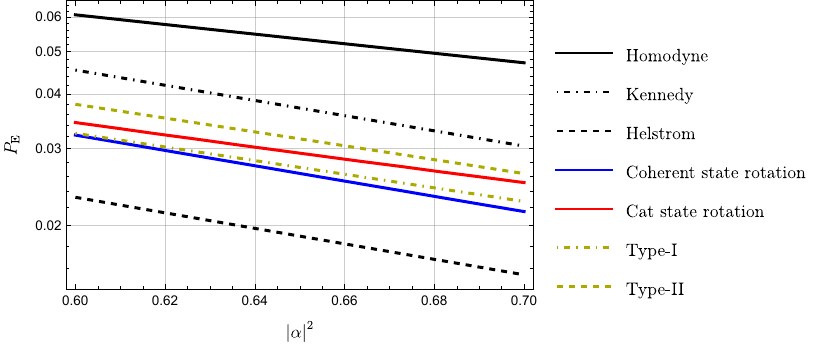} 
    \caption{Performance of the coherent state and cat state rotation-based receivers compared with the type-I and type-II near-optimal photon detection-based receivers in \cite{PhysRevA.78.022320} over the energy range $0.6<\abs{\alpha}^2<0.7$.}%
    \label{no2}%
\end{figure}

We plot comparisons in Figs.~\ref{no1} and \ref{no2}. We find that our continuously labelled receivers maintain an advantage over the photon detection based receivers for a moderate range of coherent state amplitudes. Moreover, the improved performance is achieved in the low energy limit, where the discrimination of the coherent states yields higher error rates.

\bibliographystyle{apsrev4-2-titles}
\bibliography{CVdiscBibDesk.bib}

\end{document}